\documentclass[epsfig,12pt]{article}
\usepackage{epsfig}
\usepackage{graphicx}
\usepackage{hyperref}

\usepackage{array}
\usepackage{amsmath}
\usepackage{amssymb}

\newcommand{\beq}{\begin{equation}}   
\newcommand{\eeq}{\end{equation}}
\newcommand{\beqn}{\begin{eqnarray}}   
\newcommand{\eeqn}{\end{eqnarray}}

\begin{document}
\unitlength = 1mm

\def\de{\partial}
\def\Tr{ \hbox{\rm Tr}}
\def\const{\hbox {\rm const.}}  
\def\o{\over}
\def\im{\hbox{\rm Im}}
\def\re{\hbox{\rm Re}}
\def\bra{\langle}\def\ket{\rangle}
\def\Arg{\hbox {\rm Arg}}
\def\Re{\hbox {\rm Re}}
\def\Im{\hbox {\rm Im}}
\def\diag{\hbox{\rm diag}}


\def\QATOPD#1#2#3#4{{#3 \atopwithdelims#1#2 #4}}
\def\stackunder#1#2{\mathrel{\mathop{#2}\limits_{#1}}}
\def\stackreb#1#2{\mathrel{\mathop{#2}\limits_{#1}}}
\def\Tr{{\rm Tr}}
\def\res{{\rm res}}
\def\Bf#1{\mbox{\boldmath $#1$}}
\def\balpha{{\Bf\alpha}}
\def\bbeta{{\Bf\beta}}
\def\bgamma{{\Bf\gamma}}
\def\bnu{{\Bf\nu}}
\def\bmu{{\Bf\mu}}
\def\bphi{{\Bf\phi}}
\def\bPhi{{\Bf\Phi}}
\def\bomega{{\Bf\omega}}
\def\blambda{{\Bf\lambda}}
\def\brho{{\Bf\rho}}
\def\bsigma{{\bfit\sigma}}
\def\bxi{{\Bf\xi}}
\def\bbeta{{\Bf\eta}}
\def\d{\partial}
\def\der#1#2{\frac{\d{#1}}{\d{#2}}}
\def\Im{{\rm Im}}
\def\Re{{\rm Re}}
\def\rank{{\rm rank}}
\def\diag{{\rm diag}}
\def\2{{1\over 2}}
\def\ntwo{${\mathcal N}=2\;$}
\def\nfour{${\mathcal N}=4\;$}
\def\none{${\mathcal N}=1\;$}
\def\ntwot{${\mathcal N}=(2,2)\;$}
\def\ntwoo{${\mathcal N}=(0,2)\;$}
\def\x{\stackrel{\otimes}{,}}

\newcommand{\cp}{$\mathbb{CP\,\,}$}
\newcommand{\wcp}{$\mathbb{WCP}$}

\newcommand{\cpn}{CP$(N-1)\;$}
\newcommand{\wcpn}{wCP$_{N,\widetilde{N}}(N_f-1)\;$}
\newcommand{\wcpd}{wCP$_{\widetilde{N},N}(N_f-1)\;$}
\newcommand{\vp}{\varphi}
\newcommand{\pt}{\partial}
\newcommand{\tN}{\widetilde{N}}
\newcommand{\ve}{\varepsilon}
\renewcommand{\theequation}{\thesection.\arabic{equation}}

\newcommand{\sun}{SU$(N)\;$}

\setcounter{footnote}0

\vfill

\begin{titlepage}

\begin{flushright}
FTPI-MINN-19/13, UMN-TH-3822/19\\
\end{flushright}

\begin{center}
{  \Large \bf  
 Quantizing a solitonic string
}

\vspace{5mm}

{\large \bf   M.~Shifman$^{\,a}$ and \bf A.~Yung$^{\,\,b,a,c}$}
\end {center}

\begin{center}

$^a${\it  William I. Fine Theoretical Physics Institute,
University of Minnesota,
Minneapolis, MN 55455, USA}\\
$^{b}${\it National Research Center ``Kurchatov Institute'', 
Petersburg Nuclear Physics Institute, Gatchina, St. Petersburg
188300, Russia}\\
$^{c}${\it  St. Petersburg State University,
 Universitetskaya nab., St. Petersburg 199034, Russia}
\end{center}

\begin{center}
{\large\bf Abstract}
\end{center}

Quite often the zero mode dynamics on solitonic vortices are described by
a non-conformal effective world-sheet sigma model (WSSM). We address the problem 
of solitonic string quantization in this case. As well-known, only critical strings with conformal 
WSSMs are self-consistent in ultra-violet (UV) domain.
Thus, we look for the appropriate
UV completion of the low-energy non-conformal WSSM. We argue that for the solitonic strings supported 
in  well-defined bulk theories the UV complete WSSM has a UV fixed point which can be used 
for string quantization. As an example, we consider BPS non-Abelian vortices supported by four-dimensional (4D)
\ntwo SQCD with the gauge group U$(N)$ and $N_f$ quark multiplets where $N_f\ge N$. In addition to 
translational moduli the non-Abelian vortex  under consideration carries orientational and size moduli. 
Their low-energy dynamics are described by a two-dimensional \ntwot supersymmetric weighted 
model, namely, $\mathbb{WCP}(N, N_f-N)$.
Given our  UV completion of this WSSM we find its UV fixed point. 
The latter defines a superconformal WSSM.   
We observe two cases in which this conformal WSSM, combined with 
the free theory for four translational moduli, has ten-dimensional target space required 
for superstrings to be critical.

\vspace{2cm}

\end{titlepage}



\newpage

\section {Introduction }
\label{intro}
\setcounter{equation}{0}

String theory vacua are associated with conformal two-dimensional (2D) sigma models (SMs) on the string
world sheet. This SM defines a  vacuum for the critical string theory if its Virasoro central charge equals 26 
for the bosonic string or 15 for the superstring, 
 see
for example, textbook \cite{GSW}. If not, the Liouville field does not decouple \cite{P81}
and its central charge adds up to make the total central charge with ghosts included to be zero. 

What about solitonic strings? In particular, what  can we say about confining solitonic
strings present in certain  four-dimensional (4D) gauge theories? Their quantization is a major  problem --
if resolved the solution would
give us  a first-principle framework for studying  hadronic physics.

This problem was first addressed by Polchinski and Strominger \cite{PolchStrom} for the Abrikosov-Nielsen-Olesen
(ANO) vortex \cite{ANO} in 4D Abelian-Higgs model. For the ANO vortex, the 
effective theory on the string world sheet 
reduces to the Nambu-Goto action for the translational zero modes, which in turn reduces to a free theory in the Polyakov
formulation \cite{P81} and, therefore, is obviously conformal. However, it is not critical in 4D.
The authors of \cite{PolchStrom} argued that  higher-derivative corrections improve the ultra-violet (UV)
 behavior of the theory. In particular, a six-derivative term was suggested in \cite{PolchStrom}
which is in fact the Liouville action expressed in  terms of the induced metric. 

Many solitonic strings present an even more challenging problem: their effective world sheet theory is {\em not} 
conformal. In this paper we consider (as an example) BPS non-Abelian vortices supported 
in  4D \ntwo supersymmetric QCD (SQCD) with the gauge group U$(N)$ and $N_f\ge N$, $N_f < 2N$ quark flavors.
Besides four translational moduli, the  non-Abelian vortex  have  orientational  and size moduli. 
Their low-energy dynamics is described by 2D \ntwot  weighted  \cp   
model ($\mathbb{WCP}(N,\,\tilde{N})$)
on the string world sheet \cite{HT1,ABEKY,SYmon,HT2}, with $\tN =N_f-N$  (see \cite{Trev,Jrev,SYrev,Trev2} for reviews). 
This model is not  conformal for $N_f < 2N$ and is unsuitable
 for string quantization. It has no world-sheet reparametrization invariance.

This question is puzzling. Say,  \ntwo SQCD is a completely well-defined  self-consistent theory at all distances. How come we
cannot  construct a string theory free of pathology for a solitonic vortex string supported by the above 
4D theory?

In this paper we build on the Polchinski-Strominger idea  \cite{PolchStrom}  that 
higher derivative corrections should improve UV behavior of the string world-sheet sigma model (WSSM). We will search for 
 the UV completion of the infra-red (IR)  WSSM. For a well-defined 4D theory  the UV completion of our IR WSSM should have a conformal  
fixed point in the UV which defines a quantizable string theory for the solitonic vortex. The string states of 
the UV complete theory should describe hadrons of the original 4D gauge theory.

In this paper we suggest a desired UV 
completion for the non-Abelian vortex in \ntwo SQCD. Starting from  $\mathbb{WCP}(N,\tN)$ model in the IR we find a UV-complete WSSM
 which satisfies all symmetry requirements. In particular, \ntwot supersymmetry is most restrictive.  
 
 We find two cases in which the UV fixed point becomes a 2D conformal theory (CFT)
with a 10D target space required for a superstring to be critical. 
The target space
is of the form $\mathbb{R}^4\times Y_6$, where $\mathbb{R}^4$ stands for the flat space of our 4D 
SQCD and comes from
four translational zero modes on the vortex, while $Y_6$ is a {\em non}-compact
Calabi-Yau manifold.

The first case is \ntwo  SQCD with the gauge group 
U$(N=2)$ and $N_f=3$ quark flavors. In this case $Y_6$ is the  six dimensional 
conifold, see \cite{NVafa} for a review. The infrared WSSM is asymptotically free.

The string theory on the conifold was studied previously in our papers 
\cite{SYcstring,KSYconifold,SYlittles,SYlittmult}. There,  we considered 
the non-Abelian
vortex in \ntwo  SQCD with the gauge group U$(N=2)$ and {\em four} quark flavors, $N_f=4$.
Then
the infrared WSSM is already conformal and critical and defines a string theory
for a particular value of the coupling constant where the vortex is conjectured to become infinitely thin.
The string theory on the conifold was reduced in a certain limit to a noncritical little string theory (LST)
(see \cite{Kutasov} for a review). The  spectrum of the closed string states with the lowest spins was exactly found in
\cite{SYlittles,SYlittmult}.

In this paper we do not assume that the non-Abelian vortex is infinitely thin. Instead, 
the thickness of the vortex sets a scale which plays a crucial role in our construction of the UV
complete WSSM. However, it turns out that the UV completion we suggest for  the world sheet
theory  on the non-Abelian vortex in \ntwo  SQCD with  $N_f=3$ flavors leads exactly to the string theory on the conifold mentioned above. Therefore, we 
use the results obtained in \cite{KSYconifold,SYlittles,SYlittmult} to describe the hadron spectrum
for \ntwo SQCD with $N=2$ and $N_f=3$. 

The second case in which our UV completion leads to  a critical superstring is \ntwo SQCD  with the gauge group 
U$(N=3)$ and $N_f=N=3$ quark flavors. In this case the infrared sigma model on the non-Abelian vortex is $\mathbb{CP}(N-1=2)$
model. Its UV completion has a UV fixed point with a target space  described by a non-compact Calabi-Yau manifold $Y_6$
which is the ${\cal O}(-3)$ line bundle over $\mathbb{CP}(2)$ and has local $\mathbb{CP}(2)\times\mathbb{C}$  geometry,
 see \cite{NVafa} for a review.

This case is new; the detailed study of the associated string theory is left for future work.
However, in much the same way as for the conifold case 
the string spectrum does not contain massless 4D graviton due to non-compactness of the ``extra-dimensional'' part of the target manifold $Y_6$, cf. \cite{KSYconifold,SYlittles}. This is of course a desired 
result since our starting point is 4D \ntwo SQCD without gravity.

The paper is organized as follows. In Sec.~\ref{quest} we describe general requirements for constructing the UV
completion of the IR WSSMs. In Sec.~\ref{nastrings} we review non-Abelian vortices in 
\ntwo SQCD and in particular, describe the IR $\mathbb{WCP}(N,\tN)$ models on the string world sheet. In 
Sec.~\ref{UVcompletion}
we construct UV completion of the world sheet theory and describe its UV fixed point.
In Sec. 5 we consider the critical string for  the case $N=2$ and $N_f=3$ and review our results for the 
string spectrum obtained in \cite{KSYconifold,SYlittles,SYlittmult} for the conifold case.
 In Sec. 6 we discuss the critical string for  the case $N=3$ and $N_f=3$ 
and briefly comment on expected general properties of the resulting string theory. 
Section 7 summarizes our conclusions.

\section {Quest for the UV completion}
\label{quest}
\setcounter{equation}{0}

Schematically, the world sheet theory for a solitonic vortex in 4D Yang-Mills theory can be written as 
\beq
S=\int d^2 \sigma \sqrt{h}\left\{ {\rm IR\; sigma\; model} + {\rm higher\; derivative\; terms}\right\}
\label{genaction},
\eeq
 where $\sigma^{\alpha}$ ($\alpha=1,2$) are the world-sheet coordinates, $h={\rm det}(h_{\alpha\beta})$, 
where $h_{\alpha\beta}$ is the world-sheet metric understood as an
independent variable in the Polyakov formulation \cite{P81}. The  ${\rm IR\; sigma\; model}$ has the low-energy 
sigma model action with no more than two derivatives which includes zero modes of the vortex promoted to
2D fields. 

Higher derivative corrections run in powers of the ratio 
${\pt^2}/{m_G^2}$
where $m_G$ is the scale of masses of the 4D fields which form the vortex solution. If the 4D 
theory is in the Higgs regime then  $m_G$ is the mass of the gauge and Higgs 
fields \footnote{For BPS vortex these masses are the 
same by supersymmetry.}. It determines the inverse thickness of the vortex and plays the role of the UV cutoff
for IR WSSM. At weak coupling  $m_G$ is  given by 
\beq
m_G \sim g\sqrt{T},
\label{mG}
\eeq
where $T$ is the  string tension and $g$ is a gauge coupling.

While the  IR WSSM is known in most cases and can be derived from the 4D gauge theory under consideration, 
the infinite series of higher-derivative corrections are generally unknown. Still, as we argue in Sec. \ref{intro},
they are 
important for the formulation of a well-defined string theory for a given vortex. As usual, in effective 
theories we can think that higher-derivative corrections appear as a result of integrating out  massive 
fields  residing on the string. Since higher-derivative corrections are determined by  the 4D mass $m_G$ we expect 
that these world-sheet fields have masses $ \ge m_G$. One example of such massive mode is the 
transverse size of the string itself promoted to a 2D field depending on world sheet coordinates. 

Our strategy to find a UV completion for the IR WSSM will be as follows. Instead of attempting to find 
an infinite series of higher derivative-corrections we include in the world sheet-theory massive states with mass
 $\geq m_G$. At first sight this task looks hopeless since we have to determine way 
too many massive modes. Fortunately, this is not the case. In fact, we are interested only in discrete normalizable 
modes localized near our string. Non-normalizable modes such as the  continuous spectrum of modes with the plane-wave 
asymptotics have nothing to do with the string -- they describe 
perturbative excitations present in the bulk of our 4D theory. 

Thus our task is to find a few normalizable massive modes, the mode associated with the string transverse
size being the first priority. In principle this can be done by an honest calculation, however, in this paper we 
will conjecture UV completions for the BPS non-Abelian vortices in \ntwo SQCD using the following general 
requirements.

(i) The UV completion should have a UV fixed point.

(ii) It should be \ntwot supersymmetric for the BPS vortex in \ntwo supersymmetric 4D theory.
This  is the most restrictive requirement.

(iii) The UV completion of the world sheet theory cannot have extra global symmetries not present in 
4D theory (and  in the infrared WSSM). In fact, we found that this requirement is too restrictive and 
did not allow us to find any reasonable UV completion. Therefore we replace it with a somewhat relaxed version
which is still physically reasonable.

\vspace{1mm}

(iii$^{\rm relaxed}$) If the UV completion has an additional global  symmetry not present in our 4D theory the string states
have all to be singlets  with respect to this  symmetry. Then it becomes a ``phantom'' symmetry.

This relaxed version is minimally necessary.  Indeed, the hadronic states made of strings cannot be charged with respect 
to a symmetry absent in the underlying 4D SQCD.

To conclude this section we address the following problem. One may worry that looking for the UV completion of 
the WSSM we go to high enough energies at which the vortex string can emit states living in the bulk of the 4D SQCD.
In fact, this is exactly what happens even at low energy in SQCD with $N_f>N$. As will be reviewed in the next section this theory has a Higgs branch formed by perturbative  massless states (bifundamental quarks, see next section). They can be thought of as  ``$\pi$ mesons'' in our 4D SQCD. Clearly the vortex string can emit these ``$\pi$-mesons,''
so one may worry that the problem under consideration is essentially four-dimensional. Then WSSM cannot adequately describe physics.

We  suggest that the solution of this problem can be divided in two stages. At  stage I we ignore  interactions with the 4D bulk states and  search for a well-defined WSSM {\em per se}.  The results of stage I are reprorted in this paper.The interrelation between the 2D WSSM and the given bulk theory is postponed for stage II explorations.  This
allows us to quantize the vortex string {\em per se} and find the spectrum of the string states which we interpret as hadrons of 4D SQCD.

At the second stage we must study interactions of these hadrons (string states) with the aforementioned ``$\pi$-mesons'' using an
effective Lagrangian description with interaction vertices similar to pion-nucleon vertices in QCD. 

If the ``$\pi$-meson'' energy
is small the ``$\pi$-meson'' does not probe the internal structure of a hadron. Hadron interacts as a whole and these interactions
can be described by an effective Lagrangian. The parameter of this approximation is the product of the ``$\pi$-meson'' energy and the size of the hadron which we assume to be small.

Moreover, if we include in the analysis massive states with the mass of the order of $m_G$ the problem at  stage II becomes even more complicated. In addition to interaction with ``$\pi$-mesons'' we have to take into account interactions with
massive 4D states. This is similar to nucleon interactions with $\rho$ mesons in QCD.

To summarize, at the second stage we suggest to use the same strategy that is used in effective low-energy descriptions in QCD.

In this paper we limit ourselves to the first stage. Studies of  interactions of string states with 4D
perturbative states (stage II) are left for a future work.

It is also worth noting here that the notion of UV completion of WSSM we use in this paper should be understood with care. Clearly the string formation and confinement are IR problems rather then UV problems. At very high 
energies we have quarks and gauge bosons rather than strings and hadrons. The UV completion here assumes
energies which are still lower than a certain scale $M_{\rm UV}^{2D}$, which can be thought of as a deconfinement scale.
This scale is the true UV cutoff for the WSSM.

\section {Non-Abelian vortices}
\label{nastrings}
\setcounter{equation}{0}

\subsection{Four-dimensional \boldmath{${\mathcal N}=2\;$} SQCD}

Non-Abelian vortex-strings were first found in 4D
\ntwo SQCD with the gauge group U$(N)$ and $N_f \ge N$ quark flavors  
supplemented by the FI $D$ term $\xi$
\cite{HT1,ABEKY,SYmon,HT2}, see for example \cite{SYrev} for a detailed review of this theory.
In particular, the matter sector of  the U$(N)$ theory contains
 $N_f$ quark hypermultiplets  each consisting
of   the complex scalar fields
$q^{kA}$ and $\widetilde{q}_{Ak}$ (squarks) and
their  fermion superpartners -- all in the fundamental representation of 
the SU$(N)$ gauge group.
Here $k=1,..., N$ is the color index
while $A$ is the flavor index, $A=1,..., N_f$. In this paper we assume the quark mass
parameters to vanish.
In addition, we introduce the
FI parameter $\xi$ in the U(1) factor of the gauge group.
It does not break \ntwo supersymmetry.

At weak coupling, $g^2\ll 1$ (here $g^2$ is the SU$(N)$ gauge coupling), this theory is in the Higgs regime in which squarks develop vacuum 
expectation values (VEVs).
The squark vacuum expectation values (VEV's)  are  
\beqn
\langle q^{kA}\rangle &=& \sqrt{\xi}\,
\left(
\begin{array}{cccccc}
1 & \ldots & 0 & 0 & \ldots & 0\\
\ldots & \ldots & \ldots  & \ldots & \ldots & \ldots\\
0 & \ldots & 1 & 0 & \ldots & 0\\
\end{array}
\right), \qquad  \langle\bar{\widetilde{q}}^{kA}\rangle= 0,
\nonumber\\[4mm]
k&=&1,..., N\,,\qquad A=1,...,N_f\, ,
\label{qvev}
\eeqn
where we present the squark fields as matrices in the color ($k$) and flavor ($A$) indices.

These VEVs break 
the U$(N)$ gauge group. As a result, all gauge bosons are
Higgsed. The Higgsed gauge bosons combine with the screened quarks to form 
long \ntwo multiplets with the mass 
\beq
m_G \sim g\sqrt{\xi}.
\label{ximG}
\eeq

In addition to  the U$(N)$ gauge symmetry, the squark condensate (\ref{qvev}) 
breaks also the flavor SU$(N_f)$ symmetry.
A diagonal global SU$(N)$ combining the gauge SU$(N)$ and an
SU$(N)$ subgroup of the flavor SU$(N_f)$
group survives, however.  This is a well known phenomenon of color-flavor locking. 

Thus, the unbroken global symmetry of our4D SQCD is  
\beq
 {\rm SU}(N)_{C+F}\times {\rm SU}(\tN)\times {\rm U}(1)_B,
\label{c+f}
\eeq
see \cite{SYrev} for more details. Above,  $\tN=N_f-N$. 

The unbroken global U(1)$_B$ factor in Eq. (\ref{c+f})  is identified with a baryonic symmetry. Note that 
what is usually identified as the baryonic U(1) charge is a part of  our 4D theory  gauge group.
 ``Our'' U(1)$_B$
is  an unbroken (by squark VEVs) combination of two U(1) symmetries:  the first is a subgroup of the flavor 
SU$(N_f)$ and the second is the global U(1) subgroup of U$(N)$ gauge symmetry.

The 4D theory has a Higgs branch ${\cal H}$ formed by massless quarks which are in  the bifundamental representation
of the global group \eqref{c+f} and carry baryonic charge, see \cite{KSYconifold} for more details.
The dimension of this branch is 
\beq
{\rm dim}\,{\cal H}= 4N \tN.
\label{dimH}
\eeq
 
At large $\xi$ the theory is  weakly coupled. Namely, the gauge coupling freezes at the scale $m_G$ 
(see \eqref{ximG}) and at $m_G\gg \Lambda$ we have  
\beq
\frac{8\pi^2}{g^2 (m_G)} =
(N-\tN )\ln{\frac{m_G}{\Lambda}}\gg 1 \,,      
\label{4coupling}
\eeq
were $\Lambda$ is the dynamical scale of the 4D SU($N$) gauge theory. 

\begin{figure}
\epsfxsize=10cm
\centerline{\epsfbox{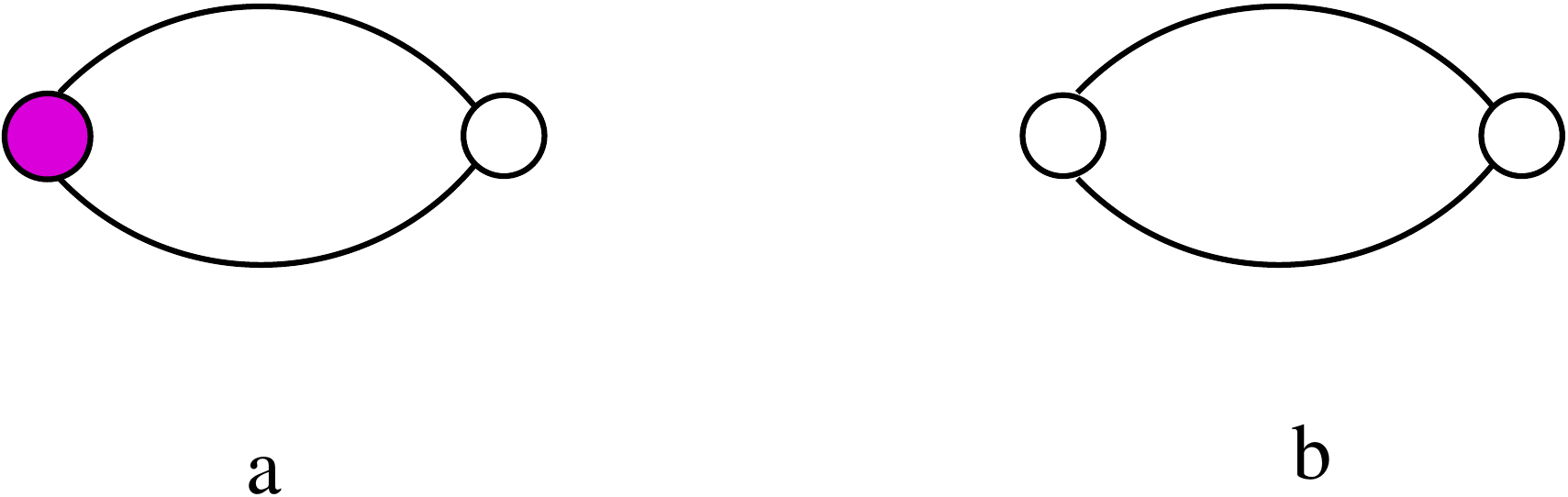}}
\caption{\small  (a) Monopole-antimonopole stringy meson. (b)  Monopole ``necklace'' baryon for U(2) gauge theory. 
Open and closed circles denote   monopoles and antimonopole respectively. }
\label{mesbaryons}
\end{figure}

As was already noted, we consider \ntwo SQCD  in the Higgs phase:  $N$ squarks  condense. 
Therefore, the non-Abelian 
vortex strings at hand confine monopoles. In the \ntwo bulk theory the above strings are 1/2 BPS-saturated; hence,  their
tension  is determined  exactly by the FI parameter,
\beq
T=2\pi \xi\,.
\label{ten}
\eeq
However, 
the monopoles cannot be attached to the string endpoints. In fact, in the U$(N)$ theories confined  
 monopoles 
are  junctions of two distinct elementary non-Abelian strings \cite{T,SYmon,HT2} (see \cite{SYrev} 
for a review). As a result,
in  4D \ntwo SQCD we have 
monopole-antimonopole mesons in which monopole and antimonopole are connected by two confining strings.
 In addition, in the U$(N)$  gauge theory we can have baryons  appearing as  a closed 
``necklace'' configurations of $N\times$(integer) monopoles \cite{SYrev}. For the U(2) gauge group the 
lightest baryon presented by such a ``necklace'' configuration  
consists of two monopoles, see Fig.~\ref{mesbaryons}.

Both stringy monopole-antimonopole mesons and monopole baryons with spins $J\sim 1$ have masses determined 
by the string tension,  $\sim \sqrt{\xi}$ and are heavier at weak coupling than perturbative states with masses
$m_G\sim g\sqrt{\xi}$. 
Thus they can decay into perturbative states \footnote{Their quantum numbers with respect to the global group 
\eqref{c+f} allow these decays, see \cite{SYrev}.} and in fact at weak coupling we do not 
expect them to appear as stable closed string states. Below  we will confirm this expectation from
the sting theory side.

If we make $\xi$ small, $\xi \ll \Lambda$ our 4D theory becomes  weakly coupled in the dual description, see
\cite{SYdualrev} for a review. The dual gauge group  U$(\tN)\times {\rm U}(1)^{N-\tN}$ is Higgsed. Vortices 
are supported in the dual theory too. They still 
confine monopoles. Quarks and gauge bosons of the original theory are in the {\em instead-of-confinement} phase
and form monopole mesons and baryons of the type shown in Fig.~\ref{mesbaryons}. For $N_f>N$ we expect that 
these states are heavy and unstable.

Only in the  ``true'' strong coupling domain $g^2\sim 1$ or $m_G \sim \Lambda$ we expect that hadrons
shown in Fig.~\ref{mesbaryons} become stable and can be described as closed string states of the soliton string theory.

\subsection{World-sheet sigma model}

The presence of the color-flavor locked group SU$(N)_{C+F}$ is the reason for the formation of the
non-Abelian vortex strings \cite{HT1,ABEKY,SYmon,HT2}.
The most important feature of these vortices is the presence of the  orientational  zero modes.
As we already mentioned, in \ntwo SQCD these strings are 1/2 BPS-saturated. 

Let us briefly review the model emerging on the world sheet
of the non-Abelian  string \cite{SYrev}.

The translational moduli fields (they decouple from all other moduli)
 in the Polyakov formulation \cite{P81} are given by the action
\beq
S_{\rm trans} = \frac{T}{2}\,\int d^2 \sigma \sqrt{h}\, 
h^{\alpha\beta}\pt_{\alpha}x^{\mu}\,\pt_{\beta}x_{\mu}
+\mbox{fermions}\,,
\label{s0}
\eeq
where  $x^{\mu}$ ($\mu=1,...,4$) 
describe the $\mathbb{R}^4$ part  of the string
target space.

If $N_f=N$  the dynamics of the orientational zero modes of the non-Abelian vortex, which become 
orientational moduli fields 
 on the world sheet, are described by two-dimensional
\ntwot supersymmetric ${\mathbb{CP}}(N-1)$ model.

If one adds additional quark flavors, non-Abelian vortices become semilocal --
they acquire size moduli \cite{AchVas}.  
In particular, for the non-Abelian semilocal vortex at hand,  in 
addition to  the complex orientational moduli  $n^P$ (here $P=1,...N$), we must add the size moduli   
$\rho^K$ (where $K=1,...\tN$), see \cite{AchVas,HT1,HT2,SYsem,Jsem,SVY}. The size moduli are also 
complex.\footnote{Both the orientational and the size moduli
have logarithmically divergent norms, see e.g.  \cite{SYsem}. After an appropriate infrared 
regularization, logarithmically divergent norms  can be absorbed into the definition of 
relevant two-dimensional fields  \cite{SYsem}. See also \cite{GKarasik} for the discussion
of different choices for quark U(1) charges which lead to a power non-normalizable orientational and  size moduli
of the vortex. In these cases orientational and  size moduli cannot be promoted to fields living on the string world sheet.}   
The low-energy dynamics of  the orientational and 
size moduli are described by the weighted \cp model, which we denote  $\mathbb{WCP}(N,\tN)$.

The  gauged formulation of  $\mathbb{WCP}(N,\tN)$ is 
as follows \cite{W93}. One introduces
 the U$(1)$ charges $\pm 1$, namely $+1$ for $n$'s and $-1$ for $\rho$'s. The bosonic part of the action reads 
\beqn
S_{\rm IR} &=& \int d^2 \sigma \sqrt{h} \left\{ h^{\alpha\beta}\left(
 \tilde{\nabla}_{\alpha}\bar{n}_P\,\nabla_{\beta} \,n^{P} 
 +\nabla_{\alpha}\bar{\rho}_K\,\tilde{\nabla}_{\beta} \,\rho^K\right)
 \right.
\nonumber\\[3mm]
&+&\left.
  2|\sigma|^2 |n^P|^2 +  2|\sigma|^2 |\rho^K|^2 + \frac{e^2}{2} \left(|n^{P}|^2-|\rho^K|^2 -\beta\right)^2
\right\}
\label{wcp}
\eeqn
where 
\beq 
\nabla_{\alpha}=\pt_{\alpha}-iA_{\alpha}\,, \qquad \tilde{\nabla}_{\alpha}=\pt_{\alpha}+iA_{\alpha},
\eeq
 while $A_\alpha$ and the complex scalar $\sigma$ form  a bosonic part of an auxiliary gauge supermultplet.
 In the limit $e^2\to\infty$ the gauged linear model (\ref{wcp}) reduces to $\mathbb{WCP}(N,\tN)$.

Classically the coupling constant $\beta$ in (\ref{wcp}) is related to 
the 4D SU(2) gauge coupling 
$g^2$ via \cite{SYrev}
\beq
\beta = \frac{4\pi}{g^2}\,.
\label{betag}
\eeq
In quantum theory the 2D coupling $\beta$ runs. The relation \eqref{betag} is 
imposed at the UV cutoff for the effective 2D theory
\eqref{wcp}. This UV cutoff is given by the scale $m_G$ which determines the inverse 
thickness of the vortex \cite{SYrev}.

Below $m_G$ the IR WSSM is asymptotically free with the coupling $\beta$  given by
\beq
\beta (\mu) =
\frac{(N-\tN )}{2\pi}\log{\frac{\mu}{\Lambda}},      
\label{2coupling}
\eeq 
where $\mu$ is the the normalization point below $m_G$.

Note that the first (and the only) coefficient of the $\beta$ functions is the same for the 4D SQCD and the
IR WSSM. This ensures that the scale of $\mathbb{WCP}(N,\tN)$ coincides with
$\Lambda$ of the 4D theory \cite{SYrev}.

The global symmetry of the IR WSSM (\ref{wcp}) 
is
\beq
 {\rm SU}(N)\times {\rm SU}(\tN)\times {\rm U}(1)_{B}\,,
\label{globgroup}
\eeq
i.e. exactly the same as the unbroken global group in the 4D theory (\ref{c+f}). 
The fields $n$ and $\rho$ 
transform in the following representations:
\beq
n:\quad (N,\,1,\, 0), \qquad \rho:\quad (1,\,\tN,\, 1)\,.
\label{repsnrho}
\eeq

Physically the  profile of a semilocal vortex in the plane orthogonal to the string axis has a two-layer 
structure. It has a hard core of radius $m_G^{-1}$ formed by heavy 4D fields and a long-range tail with power
fall-off of the profile functions at infinity. 
The tail is formed by massless quark fields fluctuating along the Higgs branch.
Moduli $\rho^K$ characterize ``sizes'' of the massless tail of the vortex \cite{AchVas,SVY}. For $N_f=N$
size moduli $\rho^K$ disappear and the model \eqref{wcp} reduces to $\mathbb{CP}(N-1)$ in the gauge formulation
\cite{W79}.

To conclude this section we note, that one can add  small masses to quarks in  4D SQCD. This will result
in adding twisted masses (equal to 4D quark masses)  to $n$ and $\rho$-fields in \eqref{wcp}, see  \cite{SYrev}. 
The twisted masses do 
not break \ntwot supersymmetry. They can be introduced by gauging a global U(1) symmetry associated with each
$n^P$ or $\rho^K$ field  and then freezing all components of the gauge multiplet, while the constant values
of the  $\sigma$ fields will determine the mass \cite{HaHo}. For simplicity we do not introduce quark masses 
in this paper.

\subsection{2D-4D correspondence}
\label{2-4}

As was mentioned above confined monopoles of 4D SQCD are junctions of two different elementary non-Abelian strings. In  the  WSSM they are seen as kinks interpolating between different vacua of
$\mathbb{WCP}(N,\tN)$ model. This ensures 2D-4D correspondence: the coincidence of the BPS spectra of 
monopoles in 4D SQCD in the quark vacuum (given by the exact Seiberg-Witten solution \cite{SW2}) and  kinks in 2D $\mathbb{WCP}(N,\tN)$ model.
This coincidence was observed in \cite{Dorey,DoHoTo} and   explained later 
in \cite{SYmon,HT2} using the picture of confined bulk monopoles which are seen as kinks in the world 
sheet theory. A crucial point is that both the monopoles and the kinks are BPS-saturated states\,\footnote{Confined
 monopoles, being junctions of two distinct 1/2-BPS strings, are 1/4-BPS states in the bulk theory 
\cite{SYmon}.},
and their masses cannot depend on the non-holomorphic parameter $\xi$ \cite{SYmon,HT2}. This means that,
although the confined monopoles look physically very different from unconfined monopoles on the Coulomb branch
of 4D SQCD,
their masses are the same. Moreover, these masses coincide with the masses of kinks in the world-sheet 
theory.

The 2D-4D correspondence imposes another very restrictive requirement on the possible UV completion of 
the IR $\mathbb{WCP}(N,\tN)$ model on the string world sheet in addition to those discussed in Sec.~\ref{quest}:

\vspace{1mm}

(iv) The UV completion should have the same spectrum of the BPS kinks as IR $\mathbb{WCP}(N,\tN)$ model
since it is fixed by 4D SQCD.

\vspace{1mm}

Below we briefly review the BPS kink spectrum in $\mathbb{WCP}(N,\tN)$ model, see \cite{SYtorkink} for  details.  It is fixed by the exact effective twisted superpotential \cite{AdDVecSal,ChVa,W93,HaHo,DoHoTo}.
Integrating out the fields $n^P$ and $\rho^K$  we obtain
 the following
exact twisted superpotential:
\beq
 {\cal W}_{\rm WCP}(\sigma)= 
\frac1{4\pi}\left\{ (N-\tN)\,
\sqrt{2}\sigma\,\ln{\frac{\sqrt{2}\sigma}{\Lambda}}
- (N-\tN) \,\sqrt{2}\sigma \right\}\, ,
\label{CPsup}
\eeq
where we use one and the same notation $\sigma$ for the  twisted superfield \cite{W93} and its lowest scalar
component. 
Minimizing this superpotential with 
respect to $\sigma$ we get the equation for the $\sigma$ VEVs  (the so-called twisted chiral ring equation),
\beq
(\sqrt{2}\sigma)^N
=\Lambda^{(N-\tN)}\,(\sqrt{2}\sigma)^{\tN}\,.
\label{sigmaeq}
\eeq
It is seen that $\tN$ roots of this equation are at $\sigma=0$ (``zero vacua'') while $(N-\tN)$ roots (``$\Lambda$-vacua'') are
\beq
\sqrt{2}\sigma = e^{\frac{2\pi i}{N-\tN} k}\,\Lambda, \qquad k=1,...(N-\tN).
\label{sigmaVEVcp}
\eeq

The  masses of the BPS kinks interpolating between two  vacua are given  by the  
differences of the superpotential (\ref{CPsup}) calculated at distinct roots \cite{HaHo,Dorey,DoHoTo},
\beq
M_{\rm BPS} =
2\left|{\cal W}_{\rm WCP}(\sigma_{1})-{\cal W}_{\rm WCP}(\sigma_{2})\right| = \frac{N-\tN}{2\pi} \Lambda\,
\left|e^{\frac{2\pi i}{N-\tN} }-1\right|.
\label{BPSmass}
\eeq
where we present the mass of the kink interpolating between  the neighboring $\Lambda$-vacua with $k=0$ and $k=1$.

If twisted masses were non-zero then the equation \eqref{sigmaeq} and the kink spectrum would become much more complicated \cite{HaHo,Dorey,DoHoTo,SYtorkink}.
In particular, due to the presence of branches  in the logarithmic functions in (\ref{CPsup}) each kink comes together with a 
tower of ``dyonic" kinks carrying global U(1) charges (for more details
see e.g. \cite{HoVa,SYtorkink}). 

The  masses obtained from (\ref{BPSmass}) were shown  
to coincide with those of the monopoles and dyons in the bulk theory. The latter are 
given by the period integrals of the Seiberg--Witten curve.

\section {The  UV completion of WSSM}
\label{UVcompletion}
\setcounter{equation}{0}

As we have already mentioned,  our WSSM in \eqref{wcp} is not conformal for $N_f <2N$ 
and cannot serve  as a sigma model for the string quantization.\footnote{The reader can keep in mind e.g. the case $N=2$, $N_f=3$, and $\tilde N =1$, see below.}
 In this section we suggest its UV completion
using requirements outlined in Sec.~\ref{quest} and Sec.~\ref{2-4}. As the simplest choice we can add a massive complex field 
$\rho_H$
with mass $\sim m_G$ to the $\mathbb{WCP}(N,\tN)$ model \eqref{wcp}, which physically describes fluctuations of the 
string hard core. We give the supermultiplet of these fields the  charge $(\tN-N)$ with respect to 
the auxiliary U(1) gauge field in \eqref{wcp}. This ensures that the associated coupling does not run at scales above $m_G$.

To preserve \ntwot supersymmetry 
and the U(1) gauge invariance while making $\rho_H$ heavy we exploit a procedure similar to that of introducing a twisted mass for
 this field  \cite{HaHo}. Namely, we gauge the global U(1) symmetry associated with $\rho_H$. To this end we introduce an extra gauge multiplet with 
 bosonic components given by gauge field $B_{\alpha}$ and a complex scalar
$\sigma_B$ assuming that $\rho_H$ has electric charge $+1$ with respect to the second gauge field.
Also, in order to avoid a second $D$-flatness condition in the UV we take the second gauge coupling $e_B^2$ finite rather than tending it to infinity
(while $e^2 \to \infty$ in \eqref{wcp}).
Moreover, to get rid of free field $\sigma_B$ in the UV we introduce a twisted superpotential 
${\cal W}^{{\rm tree}}(\sigma_B)$ which makes the second gauge multiplet heavy.

Then, the bosonic action takes the following form
\beqn
S_{\rm complete} &=& \int d^2 \sigma \sqrt{h} \left\{ h^{\alpha\beta}\left(
 \tilde{\nabla}_{\alpha}\bar{n}_P\,\nabla_{\beta} \,n^{P} 
 +\nabla_{\alpha}\bar{\rho}_K\,\tilde{\nabla}_{\beta} \,\rho^K
 +\nabla^H_{\alpha}\bar{\rho}_H\,\tilde{\nabla}^H_{\beta} \,\rho_H\right)
 \right.
\nonumber\\[3mm]
&+& \frac1{4e_B^2} B_{\alpha\beta}B^{\alpha\beta} + \frac1{e_B^2} \pt_{\alpha}\bar{\sigma}_B\pt^{\alpha}
\sigma_B  -i\sqrt{2}\, {\cal W}^{{\rm tree}}_{\sigma_B} B_{01}
\nonumber\\[3mm]
&+&
  2|\sigma|^2 |n^P|^2 +  2|\sigma|^2 |\rho^K|^2 + 2|\sigma_B -(N-\tN)\sigma |^2 |\rho_H|^2
	\nonumber\\[3mm]
&+&
\frac{e^2}{2} \left(|n^{P}|^2-|\rho^K|^2 - (N-\tN)|\rho_H|^2 
-\beta_1 \right)^2
\nonumber\\[3mm]
&+&
\left.
\frac{e_B^2}{2} \left(|\rho_H|^2 +\sqrt{2}\, {\cal W}^{{\rm tree}}_{\sigma_B}
-\beta_2 \right)^2
\right\},
\label{UVmodel}
\eeqn
where $P=1,...,N$ and  $K=1,...,\tN$, while 
\beq 
\nabla^H_{\alpha}=\pt_{\alpha}-i(N-\tN)\,A_{\alpha} +i\,B_{\alpha}\,, 
\qquad \tilde{\nabla}^H_{\alpha}=\pt_{\alpha}+i(N-\tN)\,A_{\alpha}-i\,B_{\alpha},
\eeq
$B_{\alpha\beta} $ is the field strength of the gauge field $B_{\alpha}$ and 
${\cal W}^{{\rm tree}}_{\sigma_B}$ is the  derivative of the superpotential with respect to $\sigma_B$.
We specify the exact form of the superpotential ${\cal W}^{{\rm tree}}(\sigma_B)$ later.

The subscript $H$ means ``heavy''.  The new field $\rho_H$ has an  interpretation of
a transverse size of the string core. This field enters in the action \eqref{UVmodel} on the same footing as the
``tail sizes'' $\rho^K$, namely, all $\rho$ fields have negative electric charge. The  difference is 
that $\rho_H$ is heavy with mass $\sim m_G$ as we will show below. 

 At the scales $\mu$ above the 
scale $m_G$ the 
coupling constant $\beta_1$ does not run since the sum of the electric charges
of all fields $n^P$, $\rho^K$ and $\rho_H$ is zero. The coupling constant $\beta_2$ is asymptotically free,
\beq
 \beta_{2} =
\frac1{2\pi}\log{\frac{\mu}{\Lambda_H}},     
\label{beta2}
\eeq 
where  $\Lambda_H$ is the position of the IR pole of $1/\beta_2$.
The absolute value of this scale is identified with
the mass $m_G$ in our 4D SQCD,
\beq
|\Lambda_H|= m_G,
\label{LHmG}
\eeq
while the phase is related to the $\theta$-angle of the  gauge field $B_{\alpha}$. For simplicity we assume this phase vanishing.  

Below this scale $\rho_H$ can be integrated out and the 
model \eqref{UVmodel} reduces to \eqref{wcp}. The running of its coupling is determined by 
\beq
\beta = \beta_1 +(N-\tN)\beta_2=\beta_1 + \frac{N-\tN}{2\pi}\log{\frac{\mu}{\Lambda_H}},
\label{beta1beta}
\eeq
where we use the fact that VEV of $|\rho_H|$ is given by $\beta_2$ ( ${\cal W}^{{\rm tree}}_{\sigma_B}$ is effectively zero, see below).

Comparing this with Eq.~\eqref{2coupling} we see that
\beq
 \beta_{1} =
\frac{(N-\tN )}{2\pi}\log{\frac{\Lambda_H}{\Lambda}}.     
\label{confcoupling}
\eeq 

\subsection{Exact superpotential}
\label{supUV}

The exact twisted superpotential for the model \eqref{UVmodel} is given by
\beqn
&&
 {\cal W}_{\rm eff}(\sigma, \sigma_B) = 
\frac1{4\pi}\left\{ (N-\tN)\,
\sqrt{2}\sigma\,\ln{\frac{\sqrt{2}\sigma}{\Lambda}} + 4\pi\,{\cal W}^{{\rm tree}}
\right.
\nonumber\\[3mm]
&&
\left.
+\sqrt{2}\left[\sigma_B-(N-\tN)\sigma\right]\,
\ln{\frac{\sqrt{2}\left[\sigma_B-(N-\tN)\sigma\right]}{\Lambda_H}}
- \sqrt{2}\sigma_B  \right\} ,
\label{UVsup}
\eeqn

Minimizing this superpotential with respect to $\sigma$ and $\sigma_B$ we find two vacuum equations,
namely
\beq
(N-\tN)\, \ln{\frac{\sqrt{2}\sigma}{\Lambda}}
 -(N-\tN)\,\ln{\frac{\sqrt{2}\left[\sigma_B-(N-\tN)\sigma\right]}{\Lambda_H}} =0
\label{sigmaeqUV0}
\eeq
and 
\beq
\ln{\frac{\sqrt{2}\left[\sigma_B-(N-\tN)\sigma\right]}{\Lambda_H}} + \frac{4\pi}{\sqrt{2}}\,{\cal W}^{{\rm tree}}_{\sigma_B} =0
\label{sigmaBeqUV0}.
\eeq

In Appendix A we construct  the tree superpotentials ${\cal W}^{{\rm tree}}$ for two cases, $\tN=0$ and $\tN>0$.
With this choice the vacuum equations \eqref{sigmaeqUV0} and \eqref{sigmaBeqUV0} reduce to
\beq
(\sqrt{2}\sigma)^N
=\Lambda^{(N-\tN)}\,(\sqrt{2}\sigma)^{\tN}\,,
\label{sigmaeq2}
\eeq
and 
\beq
\sqrt{2}\left[\sigma_B-(N-\tN)\sigma \right] = \Lambda_H.
\label{sigmaBvev}
\eeq
The superpotentials ${\cal W}^{{\rm tree}}$ constructed in the Appendix A satisfy the following conditions
\beq
{\cal W}^{{\rm tree}}|_{{\rm vac}} =0, \qquad {\cal W}_{\sigma_B}^{{\rm tree}}|_{{\rm vac}} =0,
\qquad \frac{\pt^2}{\pt \sigma_B^2} {\cal W}^{{\rm tree}}|_{{\rm vac}} \to \infty,
\label{supprop}
\eeq
where $|_{\rm vac}$ means that $\sigma_B$ is taken to be equal to solutions of vacuum equations
\eqref{sigmaeq2} and \eqref{sigmaBvev}.
In particular, the second condition above ensures  that the derivative of the superpotential 
${\cal W}^{{\rm tree}}_{\sigma_B}$ in Eq.~\eqref{sigmaBeqUV0}  vanish. Moreover, the third condition
makes $\sigma_B$ infinitely heavy.

Observe now that Eq.~\eqref{sigmaBvev} shows that the mass of the field $\rho_H$ is equal to $\Lambda_H$,
\beq
m_{\rho_H} = \Lambda_H,
\label{rhoHmass}
\eeq
see the third line in the Eq.~\eqref{UVmodel}. 

Eq.~\eqref{sigmaeq2} is precisely the chiral ring equation \eqref{sigmaeq} for the IR $\mathbb{WCP}(N,\tN)$ model \eqref{wcp}.
Roots of this equation are given by \eqref{sigmaVEVcp}. Calculating the mass of the BPS kink interpolating between two neighboring $\Lambda$-vacua in the model
\eqref{UVmodel} we get
\beqn
M^{\rm BPS} &=&
2\left|{\cal W}_{\rm eff}(\sigma^{(1)},\sigma_B^{(1)})-
{\cal W}_{\rm eff}(\sigma^{(2)},\sigma_B^{(2)})\right| = 
\frac1{2\pi}\left|\sqrt{2}(\sigma_B^{(1)} -\sigma_B^{(2)}\right|
\nonumber\\[3mm]
&=&\frac{N-\tN}{2\pi} \Lambda\,
\left|e^{\frac{2\pi i}{N-\tN} }-1\right|,
\label{BPSmassUV}
\eeqn
where we used Eq.~\ref{sigmaBvev}. Note, that ${\cal W}^{{\rm tree}}$ does not contribute to masses of BPS kinks due to the first condition in \eqref{supprop}.

We see that the BPS kink spectrum in our UV completion of WSSM \eqref{UVmodel}
coincides with the one \eqref{BPSmass} in the IR $\mathbb{WCP}(N,\tN)$ model.

To summarize  we outline the procedure to give a large mass $\Lambda_H$ ($|\Lambda_H|=m_G$) to the field $\rho_H$. Our procedure is similar to the standard method of introducing a twisted mass \cite{HaHo}. In the standard method the physical
degrees of freedom of extra gauge multiplet are frozen by sending $e_B$ to zero, while the VEV of $\sigma_B$ 
defines the twisted mass. Our procedure assumes that $e_B$ is finite and  we freeze the physical
degrees of freedom of the second  gauge multiplet introducing the superpotential ${\cal W}^{{\rm tree}}$. 
The advantage is that it allows us to keep the BPS kink spectrum of the IR WSSM intact. Thus we meet
a very restrictive requirement (iv), see Sec.~\ref{2-4}.

\subsection{UV fixed point}

Our WSSM \eqref{UVmodel} has a UV fixed point since the coupling $\beta_1$ does not run, while the coupling
$1/\beta_2$ is asymptotically free and goes to zero in the UV.
 The total bosonic world-sheet action is given by the sum of \eqref{UVmodel} and \eqref{s0},
\beq
S=S_{\rm trans}+S_{\rm complete}\,.
\label{stringaction}
\eeq
The UV fixed point of this \ntwot WSSM defines our superstring theory.

Since $e_B$ is finite the  $D_B$-flatness  condition does not survive in the UV. However,
the first $D$-flatness  condition in \eqref{UVmodel},  namely
\beqn
&&|n^{P}|^2-|\rho^K|^2 - (N-\tN)\,|\rho_H|^2 = \beta_1, 
\nonumber\\[2mm]
&& P=1,...,N, \quad K=1,...,\tN\,,
\label{Dflat}
\eeqn
(supplemented by factorization with respect to the U(1) gauge phase) survives in the UV and determines the 
``extra-dimensional'' target space of our string sigma model.

The above UV conformal sigma model satisfies all requirements of Sec.~\ref{quest} and Sec.~\ref{2-4}  except the condition
(iii). Clearly, adding the field $\rho_H$ increases the global symmetry of the model in the UV. For example,
if $(N-\tN) \ne 1$ we obtain an extra U(1) symmetry. Therefore, below we will use the relaxed version of the condition
(iii) and check that it is satisfied, see Sec.~\ref{quest}.

The number of real bosonic degrees of freedom in  \eqref{Dflat} is 
\beq
2(N+\tN +1 -1) = 2(N+\tN),
\eeq
where $2\times (+1)$  arises from the $\rho_H$ field , while $2\times (-1)$ is  associated with $D$-flatness 
condition and one U(1) phase eaten by the Higgs mechanism. Adding four translational moduli from
\eqref{s0} we get ten dimensional target space if 
\beq
N+\tN =3\,.
\label{critical}
\eeq
This is a condition of criticality for our superstring. 

Note, that the components (e.g. $\sigma$) of the auxiliary gauge multiplet are ''composite'' fields and do not represent independent physical degrees of freedom in the UV. In contrast, since $e_B$ is finite the components of 
$B_{\alpha}$ gauge multiplet (say $\sigma_B$) are independent degrees of freedom. We introduced superpotential 
${\cal W}^{{\rm tree}}$ to freeze  $\sigma_B$.

The condition of criticality \eqref{critical} has two solutions \footnote{ The solution with
$N=1$, $\tN =2$  gives the same theory as in \eqref{semilocal} if we take $\beta$ to be negative. In 
4D SQCD it corresponds to a dual description in the regime $m_G \ll \Lambda$.},
\beq
N=2, \qquad \tN =1
\label{semilocal}
\eeq
and
\beq
N=3, \qquad \tN =0.
\label{local}
\eeq 
In both cases the target space of the 2D sigma model has the form
\beq
\mathbb{R}^4\times Y_6, 
\label{target}
\eeq
where  $Y_6$ is a non-compact Calabi-Yau manifold.

Note that our 2D sigma model preserves ${\mathcal N} =(2,2)$ supersymmetry   which 
is a necessary condition for a superstring to have \ntwo space-time supersymmetry in 4D \cite{Gepner,BDFM}.

\section {String theory on the conifold}
\label{sec_conifold}
\setcounter{equation}{0}

In this section we will consider a critical string theory on $\mathbb{R}^4\times Y_6$ emerging in 4D SQCD with the U(2) gauge group 
and $N_f=3$ flavors. It corresponds to the first solution, Eq. \eqref{semilocal}. The electric charge of 
$\rho_H$ is $-1$ in this case so we have two $n$-fields with charge $+1$ and two $\rho$-fields with charge
$-1$. The model \eqref{UVmodel} reduces in the UV to $\mathbb{WCP}(2,2)$ model.

 As was mentioned in Sec. \ref{intro},  the string theory 
based on this sigma model  was studied earlier in our papers 
\cite{SYcstring,KSYconifold,SYlittles,SYlittmult}. 
In these papers we considered 
the non-Abelian
vortex in \ntwo  SQCD with gauge group U$(N=2)$ and four quark flavors, $N_f=4$.
 In that case
the IR WSSM was given by $\mathbb{WCP}(2,2)$. The latter model is conformal and critical 
and defines a string theory
at a particular value of the coupling constant where the vortex was conjectured to become infinitely thin.

Now we do not assume that the non-Abelian vortex is infinitely thin. Now we consider  U$(N=2)$ SQCD with
$N_f=3$, while the  additional $\rho_H$ field describes the size of the core of the non-Abelian vortex. 
However, in the 
UV limit our UV completion of the world sheet theory \eqref{UVmodel} reduces to $\mathbb{WCP}(2,2)$. Thus, we can use 
the results obtained in \cite{SYcstring,KSYconifold,SYlittles,SYlittmult} to describe the spectrum of the
closed string states in SQCD with three quark flavors. Below we review  and reinterpret these results.

Note that the global symmetry of  $\mathbb{WCP}(2,2)$ model is 
\beq
SU(2)\times U(1)_B\times SU(2)_{\rm extra}
\label{su2extra}
\eeq
so we have an extra SU(2) symmetry in our WSSM compared to the symmetry of the 4D theory,
see \eqref{c+f} for $N=2$, $\tN=1$. In this section we will see that  the string states are not charged with 
respect to this ``UV symmetry." 

The $D$-flatness condition 
takes the form
\beq
|n^{P}|^2-|\rho|^2 - |\rho_H|^2 = \beta_1, \qquad P=1,2\,,
\label{Dflatconi}
\eeq
and a U(1) phase is gauged away. This condition defines a non-compact six dimensional Calabi-Yau space, the conifold,
see \cite{NVafa} for a review.

The non-compactness is the most crucial feature of our ``extra-dimensional'' space $Y_6$. Most of the modes 
have non-normalizable wave functions over $Y_6$ and therefore do not produce dynamical fields in 4D.
Only normalizable over $Y_6$ modes localized near the tip of the conifold  can be interpreted as  hadrons 
of 4D theory. 

It is easy to see that normalizable localized states can arise only
at strong coupling in 4D SQCD. To see this we note that at weak coupling in 4D, $m_G\gg \Lambda$, according to 
\eqref{confcoupling} we have weak coupling in WSSM too, $\beta_1 \gg 1$. In this regime
the space defined by \eqref{Dflatconi} approaches a flat six dimensional space. It is clear that in this limit
there are no localized discrete states on $Y_6$. The spectrum of states is continuous, with the plane-wave asymptotics of
the wave  functions. All these states are non-normalizable. The same is true for  $m_G\ll \Lambda$ when
$\beta_1 \ll - 1$. Only at strong coupling $m_G\sim \Lambda$ or  $\beta_1\sim  0$
do we have a chance to find normalizable states.

\subsection{Massless baryon}

The only 4D massless state found in \cite{KSYconifold} is the one  associated 
with the deformation of the conifold complex structure.
 All other modes arising from the massless 10D
graviton have non-normalizable wave functions over the conifold. In particular, the 4D graviton
associated with a constant wave function over the conifold  is
absent \cite{KSYconifold}. This result matches our expectations since from the very beginning we started from
\ntwo SQCD in the flat four-dimensional space without gravity.

 Let us   construct  the U(1) gauge-invariant ``mesonic'' variables from the fields $n$ and $\rho$,
\beq
w^{PS}= n^P \rho^S.
\label{w}
\eeq
Here $\rho^S =(\rho, \rho_H)$, $S=1,2$.

These variables are subject to the constraint
${\rm det}\, w^{PS} =0$, or
\beq
\sum_{n =1}^{4} w_{n}^2 =0,
\label{coni}
\eeq
where 
$$w^{PS}\equiv \sigma_{n}^{PS}w_{n}\,,$$ 
and the $\sigma$ matrices above
are  $(1,-i\tau^a)$, $a=1,2,3$.
Equation (\ref{coni}) defines the conifold $Y_6$.  
It has the K\"ahler Ricci-flat metric and represents a non-compact
 Calabi-Yau manifold \cite{Candel,W93,NVafa}. It is a cone which can be parametrized 
by the non-compact radial coordinate 
\beq
\widetilde{r}^{\, 2} =\sum_{n =1}^{4} |w_{n}|^2\,
\label{tilder}
\eeq
and five angles, see \cite{Candel}. Its section at fixed $\widetilde{r}$ is $S_2\times S_3$.

At $\beta_1 =0$ the conifold develops a conical singularity, so both $S_2$ and $S_3$  
can shrink to zero.
The conifold singularity can be smoothed out
in two distinct ways: by deforming the K\"ahler form or by  deforming the 
complex structure. The first option is called the resolved conifold and amounts to introducing 
a non-zero $\beta_1$ in (\ref{Dflatconi}). This resolution preserves 
the K\"ahler structure and Ricci-flatness of the metric. 
If we put $\rho^K=0$ in (\ref{Dflatconi}) we get the $\mathbb{CP}(1)$ model with the $S_2$ target space
(with radius $\sqrt{\beta_1}$).  
The resolved conifold has no normalizable zero modes. 
In particular, 
the modulus $\beta_1$  which becomes a scalar field in four dimensions
 has non-normalizable wave function over the 
$Y_6$ manifold \cite{KSYconifold}.  

As  explained in \cite{GukVafaWitt,KSYconifold}, non-normalizable 4D modes can be 
interpreted as (frozen) 
parameters of the 4D  theory. 
The $\beta_1$ field is the most straightforward example of this, since the 2D coupling $\beta_1$ is
 related to the ratio $m_G/\Lambda$ in 4D SQCD, see \eqref{confcoupling}.

If $\beta_1=0$ another option exists, namely a deformation 
of the complex structure \cite{NVafa}. 
It   preserves the
K\"ahler  structure and Ricci-flatness  of the conifold and is 
usually referred to as the {\em deformed conifold}. 
It  is defined by deformation of Eq.~(\ref{coni}), namely,   
\beq
\sum_{n =1}^{4} w_{n}^2 = b\,,
\label{deformedconi}
\eeq
where $b$ is a complex number.
Now  the $S_3$ can not shrink to zero, its minimal size is 
determined by
$b$. 

The modulus $b$ becomes a 4D complex scalar field. The  effective action for  this field was calculated in \cite{KSYconifold}
using the explicit metric on the deformed conifold  \cite{Candel,Ohta,KlebStrass},
\beq
S(b) = T\int d^4x |\pt_{\mu} b|^2 \,
\log{\frac{T^2 L^4}{|b|}}\,,
\label{Sb}
\eeq
where $L$ is the  size of $\mathbb{R}^4$ introduced as an infrared regularization of 
logarithmically divergent $b$ field 
norm.\footnote{The infrared regularization
on the conifold $\widetilde{r}_{\rm max}$ translates into the size $L$ of the 4D space 
 because the variables  $\rho$ in \eqref{tilder} have an interpretation of the vortex string sizes,
$\widetilde{r}_{\rm max}\sim TL^2$ .}

We see that the norm of
the $b$ modulus turns out to be  logarithmically divergent in the infrared.
The modes with the logarithmically divergent norm are at the borderline between normalizable 
and non-normalizable modes. Usually
such states are considered as ``localized'' on the string. We follow this convention.

The field $b$,  being massless, can develop a VEV. Thus, 
we have a new Higgs branch in 4D \ntwo SQCD which opens up  only for the critical value of 
the coupling constant $\beta_1 =0$ ($m_G=\Lambda$).

 In \cite{KSYconifold} the massless state $b$ was interpreted as a baryon of 4D \ntwo SQCD.
Let us explain this.
 From Eq.~(\ref{deformedconi}) we see that the complex 
parameter $b$ (which is promoted to a 4D scalar field) is a singlet with respect to both SU(2) factors in
 (\ref{su2extra}).  What about its baryonic charge? 

Since
\beq
w_{n}= \frac12\, {\rm Tr}\left[(\bar{\sigma}_{n})_{ KP}\,n^P\rho^K\right]
\label{eq:kinkbaryon}
\eeq
we see that the $b$ state transforms as 
\beq
(1,\,2,\,1),
\label{brep}
\eeq
where we used  (\ref{repsnrho}) and (\ref{deformedconi}). 
Three numbers above refer to the representations of (\ref{su2extra}).
In particular, it has the baryon charge $Q_B(b)=2$.

As shown in \cite{KSYconifold} our string on the conifold is of type IIA.
For  type IIA superstring the complex scalar 
associated with deformations of the complex structure of the Calabi-Yau
space enters as a component of a massless 4D \ntwo hypermultiplet, see \cite{Louis} for a review. 
Instead, for type IIB superstring it would be a component of a vector BPS multiplet. Non-vanishing baryonic charge 
of the $b$ state confirms our conclusion that the string under consideration is of type IIA. The associated 
hypermultiplet
is explicitly constructed in \cite{SYlittmult}.

\subsection{Massive states}

In fact the critical string theory on the conifold is hard
to use for calculating the spectrum of massive (non-BPS) string modes because the supergravity approximation
does not work at $\beta_1=0$. In this section we review the results obtained in \cite{SYlittles} based on 
the little string theory 
(LST) approach, see \cite{SYlittles} for details. Namely,  we used the equivalent formulation of our 
string theory on the conifold 
  as a non-critical  $c=1$ string theory with the Liouville field $\phi$ and  a compact scalar $Y$ at 
the self-dual radius  
 formulated on the target space \cite{GivKut,GVafa}
\beq
\mathbb{R}^4\times \mathbb{R}_{\phi}\times S^1.
\label{targetLST}
\eeq
This  theory  
has a linear in $\phi$ dilaton, such that string coupling is given by
\beq
g_s =e^{- \frac{1}{\sqrt{2}}\phi}\, .
\label{strcoupling}
\eeq
The value of  the background charge 
of the Liouville field ($=\sqrt{2}$) ensures that the central charge of the supersymmetrized $c=1$  theory is equal to 9,
exactly what is needed for criticality.

Generically the above equivalence is formulated in a certain limit
between the critical string on the
non-compact Calabi-Yau spaces with 
an isolated singularity on the one hand, and non-critical $c=1$ string with the additional Ginzburg-Landau
\ntwo superconformal system \cite{GivKut}, on the other hand. 
In the conifold case the extra Ginzburg-Landau factor 
in \eqref{targetLST} is absent \cite{GivKutP}.

The Ginzburg-Landau superconformal system, if  present, would have a  superpotential defined by the 
left-hand side of Eq.\eqref{coni}. In this case the vertex operators would contain dependence on powers of
fields $w_{n}$ charged with respect to SU$(2)\times$ SU$(2)$ factor in \eqref{su2extra}, c.f. \cite{GivKut}.
 However, since  the Ginzburg-Landau  system is absent for the conifold case the string states are not charged 
with respect to SU(2) factors of the global group. As we will see below, they all have baryonic charge.

In fact the $c=1$ non-critical string theory  on \eqref{targetLST}  can 
also be described 
in terms of two-dimensional
black hole \cite{Wbh}, which is the ${\rm SL}(2,R)/{\rm U}(1)$ coset Wess-Zumino-Novikov-Witten theory 
\cite{MukVafa,GVafa,OoguriVafa95,GivKut} 
at level
\beq
k = 1,
\eeq
where $k$ is the total level of the Ka\v{c}-Moody algebra in the supersymmetric version (the level
of the bosonic part of the algebra is then $k_b=k+2 =3$).

In \cite{HoriKapustin} it was shown that \ntwot ${\rm SL}(2,R)/{\rm U}(1)$ coset which can be exactly solved by algebraic 
methods is a mirror description of the
$c=1$ Liouville theory.
The target 
space of 
this theory has the form of a semi-infinite cigar;  the field $\phi$ associated with the motion along the 
cigar
cannot take large negative values due to semi-infinite geometry. In this description the string
coupling constant at the tip of the cigar is $g_s \sim 1/b$. If we following \cite{GivKut} take $b$ 
large the string 
coupling at the tip of the cigar will be small and the string 
perturbation theory becomes reliable, cf. \cite{GivKut,Dorey1}. In particular,  we can use the  
tree-level approximation to obtain the string spectrum. 

The vertex operators for  the  string theory on the manifold \eqref{targetLST} are constructed in \cite{GivKut}, see also 
\cite{MukVafa,GivKutP}. Primaries of the $c=1$  part for large 
positive $\phi$ (where the target space becomes a cylinder $\mathbb{R}_{\phi}\times S^1$) take the form
\beq
V_{j,m} \sim  \exp{\left(\sqrt{2}j\phi + i\sqrt{2} m Y\right)},
\label{vertex}
\eeq
where $2m$ is integer.
Scaling dimension of the primary operator \eqref{vertex} is 
\beq
\Delta_{j,m} = m^2 - j(j+1) \, .
\label{dimV}
\eeq

The spectrum of the allowed values of $j$ and $m$ in \eqref{vertex} was  exactly determined 
 using the Ka\v{c}-Moody algebra
for the coset ${\rm SL}(2,R)/{\rm U}(1)$ in \cite{DixonPeskinLy,Petrop,Hwang,EGPerry,MukVafa}, 
see \cite{EGPerry-rev} for a review.
We will look for string states with normalizable wave functions over the ``extra dimensions'' which we will 
interpret as hadrons in 4D \ntwo SQCD. These states come from the discrete spectrum. For $k=1$  we are left 
with only two allowed values of $j$ \cite{SYlittles},
\beq
j=-\frac12, \qquad m= \pm\left\{\,\frac12,\, \frac32,...\right\}
\label{j=-1/2}
\eeq
and 
\beq
j=-1, \qquad m= \pm\{\,1, \,2,...\},
\label{j=-1}
\eeq
where $j=- 1/2$ case corresponds to the logarithmically normalizable modes like in Eq.~\eqref{Sb}.

 For scalar states in 4D the GSO projection restricts the integer $2m$ for the operator in \eqref{vertex} to be odd
\cite{KutSeib,GivKut} ,
 and we have only one possibility $j=-\frac12$, see \eqref{j=-1/2}.
This determines the masses of the 4D scalars \cite{SYlittles}, 
\beq
\frac{(M^S_{m})^2}{8\pi T} = m^2 -\frac14 \,.
\label{tachyonmass}
\eeq

In particular, the state with $m=\pm 1/2$ is the massless baryon $b$, associated with  deformations of the conifold 
complex structure \cite{SYlittles}, while states with $m = \pm (3/2, 5/2 ,...)$ are massive 4D scalars.

At the next level we consider  4D spin-2 states. 
The GSO projection selects now $2m$ to be even, $|m|=0, 1,2,...$ \cite{GivKut},
thus we are left with only one allowed value of $j$, $j=-1$ in \eqref{j=-1}. Moreover, the value
$m=0$ is excluded.
This leads to the following expression  for the masses of spin-2 states \cite{SYlittles}:
\beq
\frac{(M^{\text{spin-2}}_{m})^2}{8\pi T} = \,m^2, \qquad |m|=1,2,... .
\label{gravitonmass}
\eeq
We see that all spin-2 states are massive. This confirms the result in \cite{KSYconifold} that 
no massless 4D graviton appears in our theory. It also matches the fact that our ``boundary'' theory, 4D 
\ntwo QCD, is defined in flat space without gravity.

\begin{figure}
\epsfxsize=10cm
\centerline{\epsfbox{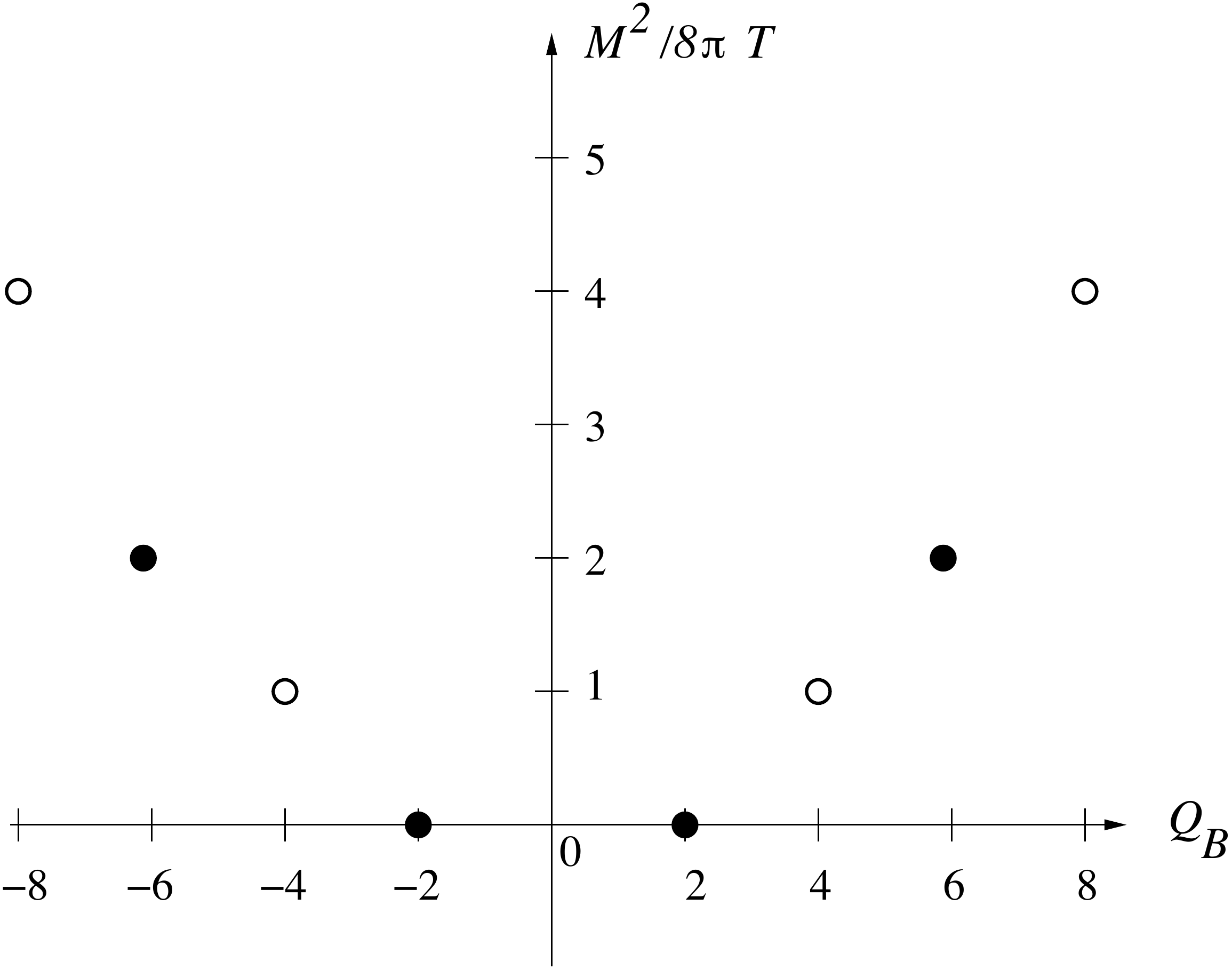}}
\caption{\small  Spectrum of spin-0 and spin-2 states as a function of the baryonic charge. Closed 
and open circles denote  spin-0 and spin-2 states, respectively.
 }
\label{fig_spectrum}
\end{figure}

The   momentum $m$ in the compact $Y$ direction of the vertex operator \eqref{vertex} is related to  the 
baryon charge of a string state \cite{SYlittles},  
\beq
 Q_B = 4m.
\label{m-baryon}
\eeq
All states reviewed above are baryons. Their  masses  as a function of the 
baryon charge are shown in Fig.~\ref{fig_spectrum}.

String states shown in \eqref{tachyonmass} and \eqref{gravitonmass} are particular representatives of \ntwo supermultiplets
in 4D. Other components can be restored by 4D supersymmetry. This was done in \cite{SYlittmult} for low-lying states.
The massless baryon in \eqref{tachyonmass} with $m=\pm 1/2$ is a hypermultiplet, while the first excited state 
with $m=\pm 3/2$ 
is a long \ntwo massive vector  supermultiplet. The lowest state with $m=\pm 1$ in \eqref{gravitonmass} contains massive 
spin-2 and vector \ntwo multiplets.

Now we can check that the condition (iii)$_{\rm relaxed}$ in Sec.~\ref{quest} is fulfilled. We see that all states 
found in \cite{SYlittles}
 have baryonic charge and, as was explained above, 
none of them are charged with respect to SU(2) factors in \eqref{su2extra}.

\section {String in U(3) SQCD}
\label{sec_U3}
\setcounter{equation}{0}

In this section we consider the string theory for the non-Abelian vortex in U(3) \ntwo SQCD with $N_f=3$ quark flavors,
see \eqref{local}. In this case $\tN=0$ and the IR WSSM does not contain $\rho$ fields at all. The 
UV completion \eqref{UVmodel} includes $\rho_H$ field with electric charge $-3$. 
In the UV limit the D-flatness
condition  reads
\beq
|n^P|^2 -3|\rho_H|^2 = \beta_1, \qquad  P=1,2,3\,,
\label{Dflat3}
\eeq
and one U(1) phase is gauged away. The coupling $\beta_1$ does not run and is determined by Eq. \eqref{confcoupling}
with $N=3$ and $\tN=0$. The sigma model target space $Y_6$ defined by \eqref{Dflat3}  is  
a non-compact Calabi-Yau manifold which is the ${\cal O}(-3)$ line bundle over $\mathbb{CP}(2)$ and locally has  
$\mathbb{CP}(2)\times\mathbb{C}$ structure, see \cite{NVafa} for 
a review. The string theory on this space 
is new and here we restrict ourselves to a few general comments leaving the detailed study of this theory for 
future work.

In much the same way as in the conifold case the  string theory on the manifold \eqref{Dflat3} does not contains 4D massless graviton.
The reason is that the manifold \eqref{Dflat3} is not compact and 4D graviton which has a constant wave function over $Y_6$
is a non-normalizable state. Of course, this conclusion match our expectations because we started with U(3) \ntwo SQCD
without gravity.

Moreover, in much the same way as in the conifold case we have a chance to find normalizable string states
only at strong coupling $m_G \sim \Lambda$ or $\beta_1 \sim 0$. To see that this is the case we note that at $|\beta_1| \to \infty$
the manifold $Y_6$ in \eqref{Dflat3} tends to flat space.

The global group of $Y_6$ in \eqref{Dflat3} is 
\beq
SU(3) \times U(1)_{\rm extra},
\label{U1extra}
\eeq
where U(1)$_{\rm extra}$ is an extra U(1) associated with the global rotation of the $\rho_H$ field. This U(1)  is absent in
4D SQCD. Below we will argue that closed string states  are not charged with respect to this 
extra U(1). 

Without the analysis of the string theory on the manifold \eqref{Dflat3}, to be carried out later, for the time being we formulate pure field theoretical
arguments. Note, that global charges of string states come from confined monopoles seen
as kinks in the world sheet theory, see \cite{SYrev} and Sec.~\ref{nastrings} above. We have shown in 
Sec.~\ref{supUV} that BPS kinks  in the model \eqref{UVmodel} coincide with kinks in the IR WSSM. Clearly
they are not charged with respect to the extra U(1) symmetry.
This implies that closed string 
states of the theory on the manifold \eqref{Dflat3} are not charged with respect to U(1)$_{\rm extra}$.  The requirement
(iii)$_{\rm extra}$ of Sec.~\ref{quest} is fulfilled.

\section {Conclusions}
\label{conclusions}
\setcounter{equation}{0}

In this paper we presented a UV completion for a conventional non-Abelian string with ${\mathbb CP}(N)$-like models on the world sheet. In our construction 
the above string flows to a conformal superstring above a certain scale $m_G$. With the judicious choice of  parameters this solitonic string becomes critical. The dependence of the string spectrum on the quark masses is not yet explored. Also, the second of two solutions presented --  the U$(N=3)$ gauge group 
 and $N_f=N=3$ quark flavors --  with the target space  described by a non-compact Calabi-Yau manifold $Y_6$ has to be further investigated. We plan to address both issues in a forthcoming publication.

\section*{Acknowledgments}

The work of M.S. is supported in part by DOE grant DE-SC0011842. 
The work of A.Y. was  supported by William I. Fine Theoretical Physics Institute,   
University of Minnesota and 
by Russian Foundation for Basic Research Grant No. 18-02-00048a.


\section*{Appendix A.  Twisted tree superpotential}

 \renewcommand{\theequation}{A.\arabic{equation}}
\setcounter{equation}{0}

In this Appendix we construct the tree twisted superpotential for our UV completion of WSSM \eqref{UVmodel},
which satisfy conditions \eqref{supprop}. Consider first the theory with $\tN=0$. We take ${\cal W}^{{\rm tree}}$
in the form
\beq
{\cal W}^{{\rm tree}} = C\,\Lambda\, \left\{\left[\frac{\sqrt{2}\sigma_B 
-\Lambda_H}{N\Lambda}\right]^{N} - 1\right\}^2,
\label{treeSup1}
\eeq
where $C$ is a constant which we take to be  large, $C\to \infty$. This will make $\sigma_B$
 infinitely heavy. The derivative of this superpotential with respect to $\sigma_B$ reads
\beq
{\cal W}^{{\rm tree}}_{\sigma_B} = 2\sqrt{2}C\, \left\{\left[\frac{\sqrt{2}\sigma_B 
-\Lambda_H}{N\Lambda}\right]^{N} - 1\right\}\left[\frac{\sqrt{2}\sigma_B 
-\Lambda_H}{N\Lambda}\right]^{N-1}.
\label{supder1}
\eeq
To check two first conditions in \eqref{supprop} we use Eq.~\eqref{sigmaBvev} to express 
$(\sqrt{2}\sigma_B  -\Lambda_H)$ in terms of $\sigma$ and then Eq.~\eqref{sigmaVEVcp} for VEVs of $\sigma$.
It is easy to see that two first conditions in \eqref{supprop} are satisfied. 

The leading contribution to the mass of $\sigma_B$ (in the limit $C\to\infty$) is proportional to the second derivative of the tree superpotential,
\beq
m_{\sigma_B} \sim e_B^2 \left|\frac{\pt^2}{\pt \sigma_B^2}{\cal W}^{{\rm tree}} \right||_{{\rm vac}}
= 4C\frac{e_B^2}{\Lambda} \left|\frac{\sqrt{2}\sigma_B -\Lambda_H}{N\Lambda}\right|^{2(N-1)}|_{{\rm vac}}
=4C\frac{e_B^2}{\Lambda}.
\eeq
We see that $\sigma_B$ becomes infinitely heavy in the limit $C\to\infty$.

Now let us consider theories with $\tN>0$. We take the tree superpotential in the form
\beq
{\cal W}^{{\rm tree}}=C\,\Lambda\,\left[\frac{\sqrt{2}\sigma_B 
-\Lambda_H}{(N-\tN)\Lambda}\right]^{2}\, 
\left\{\left[\frac{\sqrt{2}\sigma_B -\Lambda_H}{(N-\tN)\Lambda}\right]^{N-\tN}
-1\right\}^2.
\label{treeSup2}
\eeq
while its derivative reads
\beqn
&&
{\cal W}^{{\rm tree}}_{\sigma_B}=\frac{2\sqrt{2}C}{N-\tN}\,\left[\frac{\sqrt{2}\sigma_B 
-\Lambda_H}{(N-\tN)\Lambda}\right]\, 
\left\{\left[\frac{\sqrt{2}\sigma_B -\Lambda_H}{(N-\tN)\Lambda}\right]^{N-\tN}
-1\right\}
\nonumber\\[3mm]
&&
\times
\left\{(N-\tN+1)\left[\frac{\sqrt{2}\sigma_B -\Lambda_H}{(N-\tN)\Lambda}\right]^{N-\tN} -1\right\}.
\label{supderi2}
\eeqn
It is easy to see that two first conditions in \eqref{supprop} are satisfied for both  $\Lambda$ and zero-vacua.
In particular, the combination $(\sqrt{2}\sigma_B -\Lambda_H)$ is zero for zero-vacua.

Calculating the second derivative of the tree superpotential we get 
\beq
m_{\sigma_B} \sim 4C\frac{e_B^2}{(N-\tN)^2\Lambda}\,\left|(N-\tN+1)\left[\frac{\sqrt{2}\sigma_B -
\Lambda_H}{(N-\tN)\Lambda}\right]^{N-\tN} -1\right|^2|_{{\rm vac}}.
\eeq
The mass is infinite since the  absolute value above is nonzero for both  $\Lambda$ and zero-vacua.

\end{document}